\documentclass[conference]{IEEEtran}
\IEEEoverridecommandlockouts
\usepackage{cite}
\usepackage{amsmath,amssymb,amsfonts}
\usepackage{algorithmic}
\usepackage{graphicx}
\usepackage{textcomp}
\usepackage{xcolor}
\usepackage{multirow}
\usepackage[colorlinks,linkcolor=blue]{hyperref}
\usepackage{booktabs}
\def\BibTeX{{\rm B\kern-.05em{\sc i\kern-.025em b}\kern-.08em
    T\kern-.1667em\lower.7ex\hbox{E}\kern-.125emX}}
\begin{document}

\title{Contrastive Learning with Temporal Correlated Medical Images: A Case Study using Lung Segmentation in Chest X-Rays}


\author{\IEEEauthorblockN{Dewen Zeng}
\IEEEauthorblockA{\textit{University of Notre Dame}\\
Notre Dame, USA \\
dzeng2@nd.edu}
\and
\IEEEauthorblockN{John N. Kheir}
\IEEEauthorblockA{\textit{Boston Children's Hospital}\\
Boston, USA \\
John.Kheir@cardio.chboston.org}
\and
\IEEEauthorblockN{Peng Zeng}
\IEEEauthorblockA{\textit{Boston Children's Hospital}\\
Boston, USA \\
Peng.Zeng@cardio.chboston.org}
\and
\IEEEauthorblockN{Yiyu Shi}
\IEEEauthorblockA{\textit{University of Notre Dame} \\
Notre Dame, USA \\
yshi4@nd.edu}
}

\maketitle

\begin{abstract}
Contrastive learning has been proved to be a promising technique for image-level representation learning from unlabeled data.
Many existing works have demonstrated improved results by applying contrastive learning in classification and object detection tasks for either natural images or medical images.
However, its application to medical image segmentation tasks has been limited.
In this work, we use lung segmentation in chest X-rays as a case study and propose a contrastive learning framework with temporal correlated medical images, named CL-TCI, to learn superior encoders for initializing the segmentation network.
We adapt CL-TCI from two state-of-the-art contrastive learning methods-MoCo and SimCLR.
Experiment results on three chest X-ray datasets show that under two different segmentation backbones, U-Net and Deeplab-V3, CL-TCI can outperform all baselines that do not incorporate any temporal correlation in both semi-supervised learning setting and transfer learning setting with limited annotation. 
This suggests that information among temporal correlated medical images can indeed improve contrastive learning performance.
Between the two variations of CL-TCI, CL-TCI adapted from MoCo outperforms CL-TCI adapted from SimCLR in most settings, indicating that more contrastive samples can benefit the learning process and help the network learn high-quality representations. Code is available here \href{https://github.com/dewenzeng/CL-TCI}{https://github.com/dewenzeng/CL-TCI}.
\end{abstract}


\section{Introduction}

The great success of most deep learning-based models for medical image segmentation tasks highly relies on training with large-scale labeled data \cite{ronneberger2015u,xu2019whole,wang2020ica,wang2019msu,isensee2018nnu}, which is hard to get in most cases because of the requirement of expertise and extensive labeling effort.
To address this problem and achieve higher performance with limited annotation, self-supervised learning (SSL) provides a promising solution.
SSL is proposed to pre-train a model for learning general image-level features from large-scale unlabeled data without using any human-annotated labels, after which the pre-trained model can be used for finetuning on downstream tasks such as image classification \cite{he2020momentum, chen2020simple, misra2020self, grill2020bootstrap} and object detection \cite{li2020improving, xie2021detco, sun2021fsce} on label data.

Contrastive learning, a variant of SSL, has proved to be very successful in learning high-quality representations by contrasting the similarities of sample pairs in the representation space, pulling the representations of similar pairs together and pushing the representations of dissimilar pairs apart.
The typical contrastive learning frameworks include MoCo \cite{he2020momentum}, SimCLR\cite{chen2020simple} and PIRL\cite{misra2020self}, these frameworks differ in how they generate positive pairs and negative pairs as well as how they sample the data.
Existing works based on these frameworks have shown improved results in natural image classification \cite{he2020momentum, chen2020simple, misra2020self} as well as medical image classification tasks \cite{sowrirajan2020moco, vu2021medaug, azizi2021big}.

Chest X-ray (CXR) is the most common low-cost medical imaging technique for screening and identifying respiratory diseases.
One of the most important steps in computer-aided diagnosis (CAD) systems for CXR is to segment the lung and heart boundaries accurately to provide insight into cardiomegaly, pneumothorax, pneumoconiosis or emphysema \cite{carrascal1998automatic}.
Contrastive learning is a very promising technique for improving CXR interpretation models because of the large amount of unlabeled CXR data.
Despite there are some recent works such as MoCo-CXR \cite{sowrirajan2020moco} and MedAug \cite{vu2021medaug} that attempt to apply contrastive learning on large-scaled CXR dataset to improve representations for CXR interpretation, they only tested on the disease classification task, the performance of their methods on image segmentation task is still unknown.
In addition, there exist some works that use contrastive learning to improve the volumetric image segmentation model \cite{chaitanya2020contrastive, zeng2021positional}, their learning approaches depend on the characteristics of the 3D image, thus can not be applied to CXR directly.

Contrastive learning for image segmentation in CXR is different from that for natural image classification in which positive pairs are generated using different augmentations of the same image while negative pairs are views of other images.
Instead of discriminating instances in different images, the segmentation task needs to capture the features of the targets and precisely segment the targets from the background.
Besides, the target anatomical structures or organs usually exist in all images of the dataset which makes it hard to learn the ``useful'' information by conventional contrastive pair selection schema.

One important feature of the CXR data is that the same patient in Pediatric Intensive Care Unit (PICU) may have multiple CXRs taken at different timestamps so that doctors can monitor their health condition.
These CXRs are highly temporal correlated which could serve as ``good'' positive pairs with minimal mutual information \cite{tian2020makes}.
To leverage such information, we propose CL-TCI, a contrastive learning framework with temporal correlated images for the lung segmentation task in CXR.
We define CXRs coming from the same patient as positive pairs which share highly similar lung structures.
The most similar to our work is \cite{vu2021medaug} which uses metadata such as study number and laterality to generate contrastive pairs, our work differs in that we only have one laterality and the downstream task is lung segmentation.
Further, we adapt CL-TCI from both MoCo and SimCLR, named CL-TCI-MoCo and CL-TCI-SimCLR, to demonstrate the influence of data sampling on segmentation tasks.
Our contributions are:
\begin{enumerate}
    \item We develop CL-TCI to leverage the temporal correlated image of the same patient to select positive pairs in contrastive learning for lung segmentation downstream task of CXR.
    \item We demonstrate that our pre-trained model outperforms existing baselines in mean Dice for both semi-supervised setting and transfer learning setting in three different CXR datasets.
\end{enumerate}

\section{Related Work}

\noindent\textbf{Contrastive learning for Chest X-rays:}  Contrastive learning has drawn lots of attention in the medical image domain since its superior performance has been demonstrated on natural images \cite{he2020momentum,chen2020simple}.
Many existing contrastive learning works has also shown its great success on multiple tasks for CXRs, including pathology detection \cite{sowrirajan2020moco, vu2021medaug}, report generation \cite{liu2021contrastive}, and abnormality localization \cite{han2021cross}.
MoCo-CXR \cite{sowrirajan2020moco} proposed to directly apply MoCo on a large-scaled CXR dataset called CheXpert \cite{irvin2019chexpert} for pretraining and then finetuned on labeled data for diagnosis.
MedAug \cite{vu2021medaug} improved MoCo-CXR by using patients' metadata to generate positive and negative pairs.
\cite{liu2021contrastive} introduced a contrastive attention model to contrast the current input image with normal images to distill the suspicious abnormal regions for generating better CXR reports automatically.
\cite{han2021cross} presented a novel idea of generating positive pairs by contrasting two different modalities characterizing the same patient to improve abnormality identification and localization in CXR.
Despite these successes, the idea of applying contrastive learning on segmentation tasks for CXR is still unexplored.

\noindent\textbf{Contrastive learning for medical image segmentation:}
Prior self-supervised feature learning methods for medical image segmentation mostly focus on designing different pretext tasks such as Rubik's cube \cite{zhuang2019self,zhu2020rubik}, anatomical position prediction \cite{bai2019self}, patch distance prediction \cite{spitzer2018improving}, rotation and jigsaw puzzle \cite{taleb20203d}.
Recently, contrastive learning-based approach has become the state-of-the-art SSL method because it does not rely on ad-hoc heuristics which may cause less generality of the pre-trained model. \cite{chaitanya2020contrastive} proposed a global and local contrastive learning framework for volumetric medical image segmentation with limited annotation. 
PCL \cite{zeng2021positional} improved the global contrastive learning by introducing the position of the 2D slices in the volumetric image to select contrastive pairs.
Pgl \cite{xie2020pgl} proposed a prior-guided self-supervised model to learn the region-wise local consistency in the latent feature space for segmentation tasks.
All of these methods target 3D medical image segmentation tasks, assuming that the anatomical information of the corresponding location of different images is similar.
This might not work for 2D images like CXR in which the scale and position of the target in different acquisitions vary a lot.
In this work, we want to address the contrastive learning problem for segmentation tasks in CXRs.


\section{Method}

\begin{figure*}[!htb]                               
\centering
\includegraphics[width=2.0 \columnwidth]{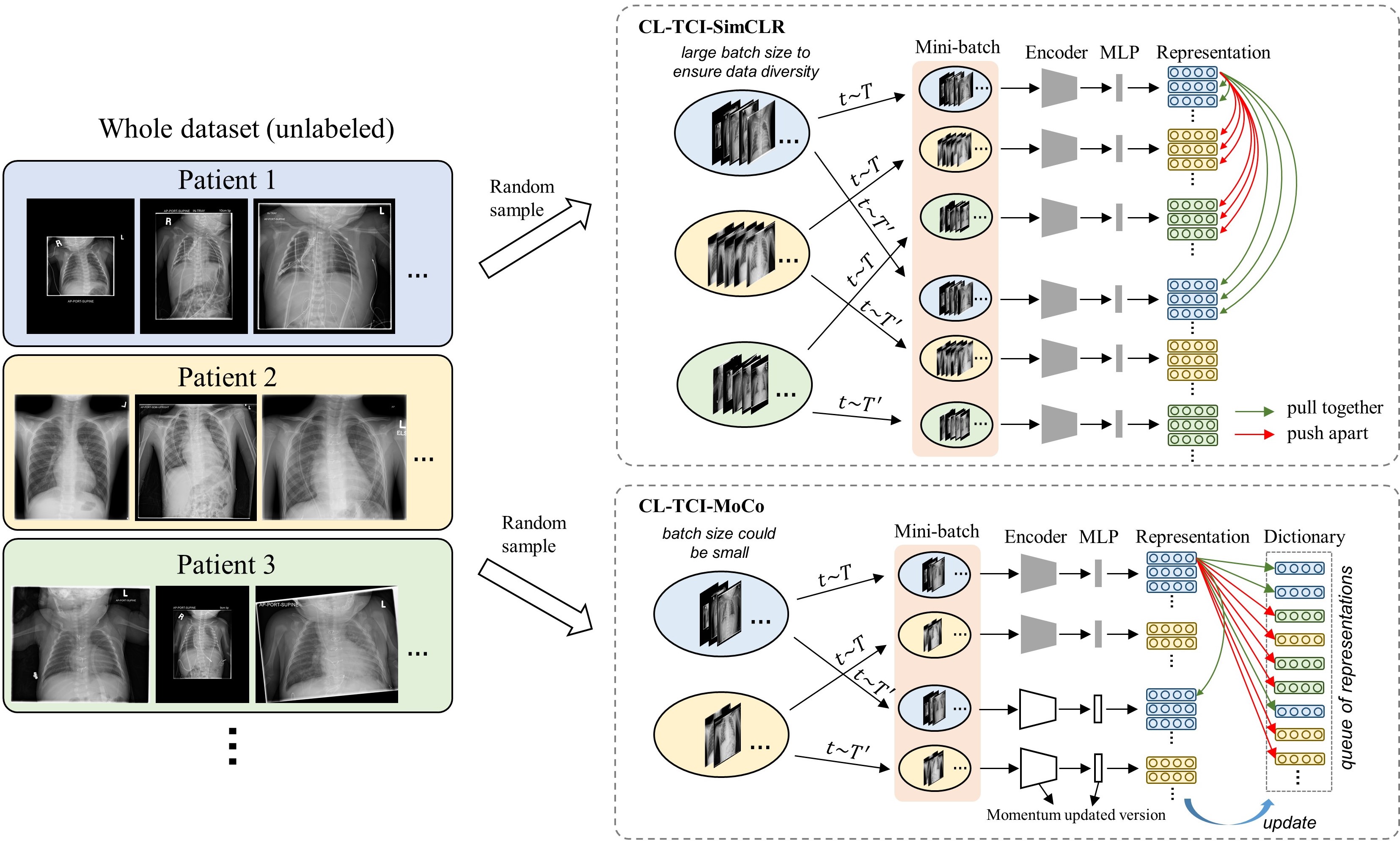}
\caption{Overview of the proposed contrastive learning framework with temporal correlated medical images (CL-TCI). Chest X-ray data are sampled from different patients and fitted into the Encoder and MLP for representation learning, one representation vector for each sample. Features from the same patient (marked as the same color) are considered positive and should be pulled together in the feature space. In contrast, features from different patients are negative and should be pushed apart. In CL-TCI-SimCLR, each representation vector contrasts with all the other vectors in the mini-batch, while in CL-TCI-MoCo each representation vector contrasts with its transformed version plus all the representations in the dictionary. }
\label{fig:method_overview}
\end{figure*}

Our contrastive learning framework with temporal correlated image can be built upon SimCLR \cite{chen2020simple} and MoCo \cite{he2020momentum}, which we called CL-TCI-SimCLR and CL-TCI-MoCo, for unsupervised CXR image representation learning.
It consists of two stages: (1) the pre-training stage (2) the fine-tuning stage.
We first discuss the pre-training stage where contrastive learning is used for representation learning in Section \ref{section:pre-training}.
Then the fine-tuning stage will be introduced in Section \ref{section:fine-tuning}.

\subsection{Pre-training with temporal correlated images}
\label{section:pre-training}

Different from most existing CXR segmentation datasets where each patient has only one image, our domestic collected dataset contains multiple CXRs collected at a different time for each patient in order to monitor the patient's conditions with time.
Although CXRs of the same patient may appear different because of the view or health conditions, the overall lung area and features are very similar.
Such information can be utilized in contrastive learning for image representation learning without any annotation.

\subsubsection{CL-TCI-SimCLR}

The overview of the proposed CL-TCI-SimCLR is shown in Fig. \ref{fig:method_overview}.
Firstly, $N$ chest X-rays are randomly sampled from different patients.
Each sampled instance is augmented using standard appearance and geometric transformation ($T$ and $T^\prime$ in Fig. \ref{fig:method_overview}) such as rotation and horizontal flip, resulting in 2$N$ X-rays.
These $2N$ instances form a mini-batch for representation learning.
Then, each transformed instance is forwarded to a CNN Encoder followed by a non-linear MLP (multi-layer perceptron) projection head to generate the representation vector.
Here, suppose $x_i$ represents the original image sampled from the patients, $x_{2i}^{\prime}$ and $x_{2i-1}^{\prime}$ are two augmented version of $x_i$.
The representation vectors generated from $x_{2i}^{\prime}$ and $x_{2i-1}^{\prime}$ are denoted as $z_{2i}$ and $z_{2i-1}$.

In traditional contrastive learning methods such MoCo \cite{he2020momentum} and SimCLR \cite{chen2020simple} for image-level representation learning, the two augmented versions $x_{2i}^{\prime}$ and $x_{2i-1}^{\prime}$ of the same sample $x_i$ are considered as positive and should be pull together in the representation space.
$x_{2i-1}^{\prime}$ of the same sample $x_i$ with all the other instances in the mini-batch or the memory bank are negative and should be pushed apart in the representation space.
InfoNCE loss \cite{oord2018representation} can be used for training the network.
However, in our contrastive learning framework, all CXRs from the same patient are considered as positive and those from different patients are negative (see Fig. \ref{fig:method_overview}, positive samples are labeled/marked by the same color).
Therefore, more than one positive pair can be present in a mini-batch.
In order to handle multiple positive samples during training, we use an extended contrastive loss which was introduced originally in SupContrast \cite{khosla2020supervised} for handling arbitrary numbers of positives. 
The loss function can be formulated as:

\begin{equation}
    \mathcal{L}=\sum_{i=1}^{2N}{\mathcal{L}_{i}},
\label{eq:1}
\end{equation}
\begin{equation}
    \mathcal{L}_{i}=-\frac{1}{|\Omega_{i}^{+}|}\sum_{j\in \Omega_{i}^{+}}{log\frac{e^{sim(z_i,z_j)/\tau}}{\sum_{k=1}^{2N}\mathbb{1}_{i\neq k}\cdot e^{sim(z_i,z_k)/\tau}}}.
\label{eq:2}
\end{equation}
where $\Omega_{i}^{+}$ is the set of indices of positive instances to $x_i^\prime$ which includes the augmented version and other instances from the same patient. 
$sim(\cdot,\cdot)$ is the cosine similarity function that computes the similarity between two vectors in the representation space. 
$\tau$ is a temperature scaling parameter.
The major advantage of this loss function is that all positive instances in the mini-batch contribute to the numerator, allowing for a robust clustering of the representation space.
In order to make sure there exist enough positive and negative pairs in each mini-batch, a large batch size needs to be used in CL-TCI-SimCLR for more data diversity.

\begin{table*}[t]
\tabcolsep 11pt
\centering
\caption{Comparison of the proposed framework in \textbf{semi-supervised setting} with baseline methods using U-Net and Deeplab V3+ as the backbone. $M$ is the number of images used in the fine-tuning process. Results are reported in the form of mean(standard deviation) on 5-fold cross-validation.
}
\begin{tabular}{clcccccc}
\toprule
Backbone & Method & $M$=10 & $M$=20 & $M$=50 & $M$=100 & $M$=150 & $M$=246 \\ \hline 
\multirow{8}{*}{U-Net} & Random & 0.771(.121) & 0.870(.010) & 0.906(.010) & 0.920(.010) & 0.926(.004) & 0.930(.004) \\ 
& Rotation~\cite{gidaris2018unsupervised} & 0.781(.040) & 0.879(.009) & 0.909(.009) & 0.920(.007) & 0.928(.005) & 0.931(.004) \\
& PIRL~\cite{misra2020self} & 0.550(.144) & 0.865(.051) & 0.904(.009) & 0.921(.005) & 0.928(.005) & 0.931(.004)  \\
& SimCLR~\cite{chen2020simple} & 0.851(.013) & 0.883(.011) & 0.909(.006) & 0.918(.005) & 0.924(.004) & 0.929(.003) \\
& MoCo~\cite{he2020momentum} & 0.846(.014) & 0.881(.012) & 0.909(.007) & 0.920(.005) & 0.925(.005) & 0.930(.002) \\
& CL-TCI-SimCLR & 0.858(.008) & 0.885(.010) & 0.911(.010) & 0.924(.004) & 0.928(.003) & 0.932(.004) \\
& CL-TCI-MoCo & \textbf{0.877(.004)} & \textbf{0.898(.005)} & \textbf{0.917(.004)} & \textbf{0.926(.004)} & \textbf{0.931(.004)} & \textbf{0.934(.004)} \\ \hline
\multirow{8}{*}{Deeplab V3+} & Random & 0.766(.017) & 0.836(.009) & 0.893(.009) & 0.913(.005) & 0.920(.006) & 0.927(.004) \\ 
& Rotation~\cite{gidaris2018unsupervised} & 0.804(.025) & 0.851(.012) & 0.892(.008) & 0.909(.007) & 0.921(.004) & 0.926(.005) \\
& PIRL~\cite{misra2020self} & 0.681(.027) & 0.768(.013) & 0.870(.006) & 0.898(.004) & 0.912(.005) & 0.920(.004) \\
& SimCLR~\cite{chen2020simple} & 0.829(.010) & 0.864(.010) & 0.901(.008) & 0.916(.004) & 0.923(.004) & 0.927(.003) \\
& MoCo~\cite{he2020momentum} & 0.849(.012) & 0.881(.009) & 0.907(.007) & 0.919(.003) & 0.925(.006) & 0.931(.003) \\
& CL-TCI-SimCLR & 0.871(.006) & 0.893(.004) & 0.915(.005) & 0.925(.003) & \textbf{0.930(.004)} & 0.932(.003) \\
& CL-TCI-MoCo & \textbf{0.877(.003)} & \textbf{0.896(.004)} & \textbf{0.916(.005)} & \textbf{0.925(.004)} & \textbf{0.930(.004)} & \textbf{0.934(.003)} \\
\bottomrule
\end{tabular}
\label{table:semi_supervised_setting}
\end{table*}

\subsubsection{CL-TCI-MoCo}

The illustration of CL-TCI-MoCo can be seen in Fig. \ref{fig:method_overview}.
Different from CL-TCI-SimCLR where the representation vector of each sample is in contrast with all the other samples inside the mini-batch, in CL-TCI-MoCo each representation vector $z_{2i}$ needs to contrast with its another transformed version $z_{2i-1}$ plus a \textit{queue} of representation maintained in a dictionary proposed in MoCo \cite{he2020momentum}.
The dictionary is dynamically updated during training and the keys are randomly sampled, the dictionary update follows the first in first out strategy and the momentum update follows the original MoCo.
For each representation inside the dictionary, we maintain a pseudo-label generated by the patient id (shown in different colors in Fig. \ref{fig:method_overview}) to record which patient this representation is generated from.
During contrastive learning, representations originated from the same patient are positive and those from different patients are negative, so the SupContrast loss discussed in CL-TCI-SimCLR can still be used.

Note that each representation $z_{2i}$ does not compute with other representations in the mini-batch except $z_{2i-1}$ in CL-TCI-MoCo, the batch size will not affect the data diversity during training while the size of the dictionary does.
Therefore, CL-TCI-MoCo is more memory and energy-efficient because the batch size can be small so that the training needs less computational resources.

\subsection{Fine-tuning for lung segmentation in chest X-ray}
\label{section:fine-tuning}


After pre-training finishes, the MLP is thrown away and the encoder is used to initialize the encoder of the segmentation network in the fine-tuning stage.
Then, the segmentation network is fine-tuned with limited annotation using supervised learning to achieve the highest possible accuracy.
In order to show the robustness of the proposed contrastive learning framework, we use two standard segmentation architectures: U-Net \cite{ronneberger2015u} and Deeplab V3+ \cite{chen2018encoder} as the backbone in the fine-tuning stage which have shown remarkable success in many medical image segmentation tasks \cite{isensee2018nnu,kohl2018probabilistic,choudhury2018segmentation}.

\section{Experiment}

\begin{table*}[t]
\tabcolsep 11pt
\centering
\caption{Transfer learning results on JSRT dataset using U-Net and Deeplab V3+ as the backbone. $M$ is the number of images used in the fine-tuning process.
}
\begin{tabular}{clcccccc}
\toprule
Backbone & Method & $M$=10 & $M$=20 & $M$=50 & $M$=100 & $M$=150 & $M$=197 \\ \hline 
\multirow{8}{*}{U-Net} & Random & 0.917(.009) & 0.950(.001) &  0.961(.002) & 0.966(.002) & 0.968(.002) & 0.969(.002) \\ 
& Rotation~\cite{gidaris2018unsupervised} & 0.909(.009) & 0.949(.001) & 0.962(.002) & 0.965(.002) & 0.967(.002) & 0.968(.002) \\
& PIRL~\cite{misra2020self} & 0.860(.030) & 0.946(.003) & 0.961(.002) & 0.966(.002) & 0.967(.002) & 0.969(.002) \\
& SimCLR~\cite{chen2020simple} & 0.938(.005) & 0.955(.001) & 0.964(.001) & 0.967(.001) & 0.968(.001) & 0.969(.002) \\
& MoCo~\cite{he2020momentum} & 0.938(.004) & 0.954(.001) & 0.963(.002) & 0.967(.001) & 0.968(.002) & 0.969(.001) \\
& CL-TCI-SimCLR & 0.938(.003) & 0.958(.003) & 0.964(.002) & 0.967(.001) & 0.968(.002) & 0.969(.001) \\
& CL-TCI-MoCo & \textbf{0.943(.003)} & \textbf{0.959(.001)} & \textbf{0.966(.001)} & \textbf{0.969(.001)} & \textbf{0.970(.001)} & \textbf{0.970(.001)} \\ \hline
\multirow{8}{*}{Deeplab V3+} & Random & 0.898(.007) & 0.933(.003) & 0.952(.002) & 0.961(.002) & 0.964(.002) & 0.965(.002)  \\ 
& Rotation~\cite{gidaris2018unsupervised} & 0.924(.007) & 0.943(.003) & 0.956(.002) & 0.962(.002) & 0.964(.002) & 0.965(.002) \\
& PIRL~\cite{misra2020self} & 0.885(.008) & 0.916(.006) & 0.942(.001) & 0.954(.001) & 0.957(.001) & 0.960(.001) \\
& SimCLR~\cite{chen2020simple} & 0.927(.005) & 0.944(.002) & 0.957(.001) & 0.963(.002) & 0.965(.002) & 0.966(.002) \\
& MoCo~\cite{he2020momentum} & 0.936(.004) & 0.954(.006) & 0.962(.002) & \textbf{0.966(.002)} & 0.967(.002) & \textbf{0.968(.002)} \\
& CL-TCI-SimCLR & 0.944(.004) & 0.955(.001) & \textbf{0.963(.002)} & \textbf{0.966(.002)} & \textbf{0.967(.001)} & \textbf{0.968(.002)} \\
& CL-TCI-MoCo & \textbf{0.945(.002)} & \textbf{0.956(.001)} & \textbf{0.963(.002)} & \textbf{0.966(.002)} & \textbf{0.967(.001)} & \textbf{0.968(.002)} \\
\bottomrule
\end{tabular}
\label{table:transfer_learning_jsrt}
\end{table*}

\begin{figure*}[!htb]                               
\centering
\includegraphics[width=2.0 \columnwidth]{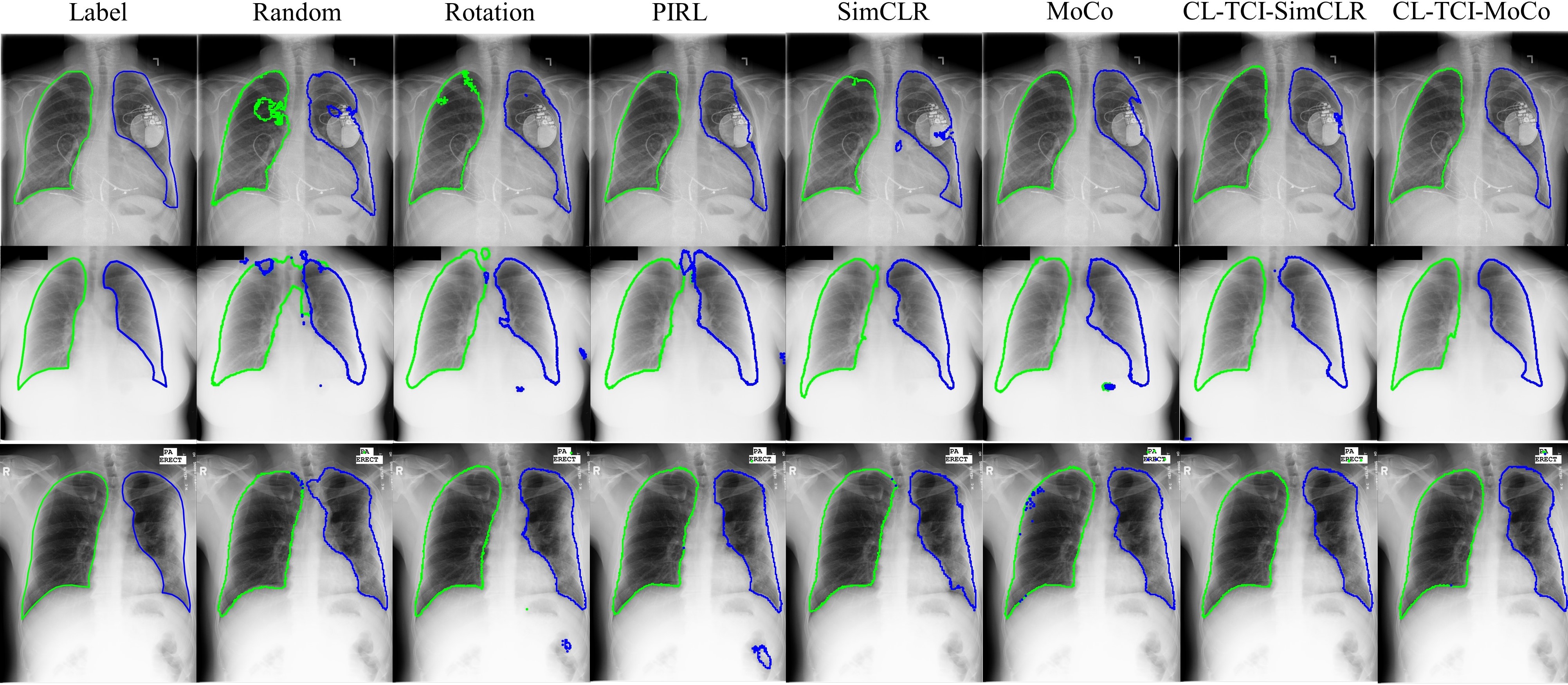}
\caption{Visualization of segmentation results on all datasets using U-Net backbone. The blue and green boundaries represent left lung and right lung, respectively. From top to bottom, each row denotes the results of BCH-CXR-subset, JSRT and Montgomery dataset. The results of BCH-subset are generated from fine-tuned model when $M=100$ while the results of JSRT and Montgomery are generated from fine-tuned models when $M=20$.}
\label{fig:visualization_results}
\end{figure*}

\subsection{Semi-supervised learning}

\label{section:semi-supervised}
In this section, we evaluate the performance of the proposed framework in the semi-supervised setting where the pre-training and fine-tuning are done on the same dataset.

\textbf{Dataset:} The chest X-rays we use to evaluate the proposed contrastive learning framework are collected from Boston Children's Hospital, Boston, USA.
The dataset contains chest X-rays from 105 patients, each patient has 1 to 120 images with diverse image dimensions from 900$\times$900 to 2400$\times$2400, the width and the height of the image are not necessarily the same.
The total number of images is 3121.
No label is given for these images, we denote this dataset as BCH-CXR.
Then, we randomly sample 308 images from BCH-CXR and ask an experienced radiologist to annotate the left lung and right lung of these images. We denote this annotated dataset as BCH-CXR-subset.
In the following experiment, BCH-CXR is used in the pre-training stage for contrastive learning. BCH-CXR-subset is used in the fine-tuning stage for supervised learning.

\textbf{Setup:} Since the dimension of the images in the dataset are diverse, we first square all images by padding 0 to the shorter axis and then scale them to 256$\times$256 pixels which retrain the visual details of the lung fields.
We employ the proposed contrastive learning framework to train a U-Net encoder and a Deeplab V3+ encoder (Resnet50) on BCH-CXR.
Then the pre-trained U-Net encoder and Resnet50 encoder are used for the initialization of the associated networks for fine-tuning on BCH-CXR-subset. 
Notice that BCH-CXR-subset is a subset of BCH-CXR, so this meets the definition of semi-supervised learning.
Five-fold cross-validation and Dice coefficient are used to evaluate the segmentation performance.
In order to investigate the framework on limited annotation, we randomly sample $M, M\in \{10,20,50,100,...\}$ in training folds as if we only have the labels for these images, and evaluate the result on the validation fold.
We form the lung segmentation problem as a three-class segmentation problem, which means the left lung and right lung are defined as different classes in the annotation.
This is necessary to avoid additional post-processing because some clinical metrics need to be calculated on the left lung and right lung separately.

We compare the performance of our proposed framework with a random approach that does not use any pre-training as well as the following state-of-the-art baselines, all of which use the same unlabeled dataset in the pre-training and labeled dataset in the fine-tuning as our proposed framework:
(1) Rotation \cite{gidaris2018unsupervised}: a pretext-based method that uses image rotation prediction to pre-train the encoder; 
(2) PIRL \cite{misra2020self}: adopted from a contrastive learning scheme for natural image classification, which uses contrastive loss to learn pretext-invariant representations.
(3) SimCLR \cite{chen2020simple}: adopted from another contrastive learning scheme for natural image classification, which constructs positive pairs for each sample only using two random augmentations.
(4) MoCo \cite{he2020momentum}: adopted from another contrastive learning scheme for natural image classification, which introduces momentum update and dictionary look-up to facilitate contrastive learning.
We implement two versions of our methods: (5) CL-TCI-SimCLR and (6) CL-TCI-MoCo.

Data augmentation including rotation, translation, and scale are used in both the pre-training and fine-tuning stages.
The pre-training is done on two NVIDIA GeForce 1080 GPUs with 500 epochs. 
SGD is used as the optimizer and the learning rate is set to 0.1 with a cosine learning schedule.
Batch size is set to 32 and 16 for SimCLR-related methods and MoCo-related methods, respectively. 
Temperature $\tau$ is set to 0.1 as in \cite{he2020momentum,chen2020simple}.
During fine-tuning, the segmentation network is trained for 200 epochs with a batch size of 10.
Adam optimizer is used and the learning rate is set to $5\times e^{-5}$ with cosine scheduler.

The quantitative results of using U-Net backbone and Deeplab V3+ backbone are shown in Table \ref{table:semi_supervised_setting}.
It can be seen that compared to all the baselines, our CL-TCI perform the best for all values of $M$, suggesting that the proposed framework can learn better representations for downstream segmentation tasks.
The gains become lesser when the number of training samples increases. 
This is because with more training samples, the information difference between the training set for fine-tuning and the training set for contrastive learning becomes small and the fine-tuning performance saturates.
Rotation can sometimes improve the accuracy when the finetuning samples are small, but the improvement is marginal.
PIRL does not outperform Random in most settings, suggesting that solving the Jigsaw puzzle might not help the network learn segmentation-related features.
SimCLR and MoCo can both benefit the segmentation performance, they behave similarly with U-Net backbone while MoCo is slightly better than SimCLR with Deeplab V3+ backbone.
Comparing CL-TCI-Moco and CL-TCI-SimCLR, we can see that CL-TCI-Moco consistently performs better, proving that leveraging more contrastive pairs in our CL-TCI framework can further improve the segmentation performance when training samples for finetuning is small.

\begin{table*}[t]
\tabcolsep 11pt
\centering
\caption{Transfer learning results on Montgomery dataset using U-Net and Deeplab V3+ as the backbone. $M$ is the number of images used in the fine-tuning process.
}
\begin{tabular}{clccccc}
\toprule
Backbone & Method & $M$=10 & $M$=20 & $M$=50 & $M$=100  \\ \hline 
\multirow{8}{*}{U-Net} & Random & 0.945(.005) & 0.964(.002) & 0.974(.002) & 0.978(.002) \\ 
& Rotation~\cite{gidaris2018unsupervised} & 0.937(.008) & 0.957(.011) & 0.972(.002) & 0.977(.002) \\
& PIRL~\cite{misra2020self} & 0.916(.017) & 0.962(.001) & 0.972(.002) & 0.977(.002) \\
& SimCLR~\cite{chen2020simple} & 0.943(.003) & 0.966(.001) & 0.974(.002) & 0.978(.002) \\
& MoCo~\cite{he2020momentum} & 0.952(.004) & 0.967(.003) & \textbf{0.975(.002)} & 0.978(.002) \\
& CL-TCI-SimCLR & 0.953(.004) & 0.967(.005) & \textbf{0.975(.002)} & 0.978(.002)  \\
& CL-TCI-MoCo & \textbf{0.957(.003)} & \textbf{0.968(.002)} & \textbf{0.975(.002)} & \textbf{0.979(.002)}  \\ \hline
\multirow{8}{*}{Deeplab V3+} & Random & 0.912(.004) & 0.941(.004) & 0.962(.002) & 0.971(.002) \\ 
& Rotation~\cite{gidaris2018unsupervised} & 0.916(.011) & 0.942(.010) & 0.962(.009) & 0.973(.002) \\
& PIRL~\cite{misra2020self} & 0.905(.008) & 0.934(.004) & 0.957(.003) & 0.966(.003)  \\
& SimCLR~\cite{chen2020simple} & 0.935(.004) & 0.953(.004) & 0.969(.002) & 0.974(.002)  \\
& MoCo~\cite{he2020momentum} & 0.947(.007) & 0.962(.003) & 0.972(.003) & 0.976(.002) \\
& CL-TCI-SimCLR & 0.947(.007) & 0.963(.002) & 0.972(.002) & 0.976(.001) \\
& CL-TCI-MoCo & \textbf{0.949(.005)} & \textbf{0.963(.002)} & \textbf{0.973(.001)} & \textbf{0.977(.002)}  \\
\bottomrule
\end{tabular}
\label{table:transfer_learning_montgomery}
\end{table*}

\subsection{Transfer learning}
In this section, we want to evaluate whether the learned representations by the proposed framework are transferrable to other datasets.
We use the encoder learned in Section \ref{section:semi-supervised} to initialize the segmentation network to fine-tune on JSRT \cite{shiraishi2000development} and Montgomery \cite{jaeger2014two} dataset.
JSRT dataset contains 247 CXRs, among which 154 have lung nodules and 93 have no lung nodule.
The image size in JSRT dataset is 2048$\times$2048 pixels. 
Montgomery contains 138 CXRs, including 80 normal patients and 58 patients with manifested tuberculosis (TB), the image size in this dataset is 4020 $\times$ 4892 or 4892 $\times$ 4020.
We resize all the images to $256\times 256$ pixels and the experiment setup and baseline are the same as the fine-tuning stage in Section \ref{section:semi-supervised}.

The comparison results on JSRT and Montgomery datasets using U-Net and Deeplab V3+ as the backbone are shown in Table \ref{table:transfer_learning_jsrt} and Table \ref{table:transfer_learning_montgomery}, respectively.
It can be seen that our CL-TCI methods still outperform all the baselines for all $M$ values. 
CL-TCI-MoCo consistently performs the best in all settings.
The gains are much less than the gains on BCH-CXR-subset, this is because the CXRs in JSRT and Montgomery are more standard and consistent in terms of acquisition position and acquisition view than the BCH-CXR dataset, making the lung segmentation task on these two datasets easier than that of on BCH-CXR-subset.
PIRL is worse than random initialization.
Rotation can sometimes improve the performance but the improvement is limited.
We can also notice that MoCo performs better than SimCLR in all transfer learning settings.
In general, the experiment results show that the representations learning by the proposed contrastive learning framework can be transferred to improve the segmentation accuracy on other unseen datasets.
Visualization of segmentation results in semi-supervised setting and transfer learning setting can be seen in Fig. \ref{fig:visualization_results}, we can see that both CL-TCI-SimCLR and CL-TCI-MoCo can produce more accurate lung boundaries than other baselines.

\subsection{Visualization of learned features}

\begin{figure}[!htb]                               
\centering
\includegraphics[width=1.0 \columnwidth]{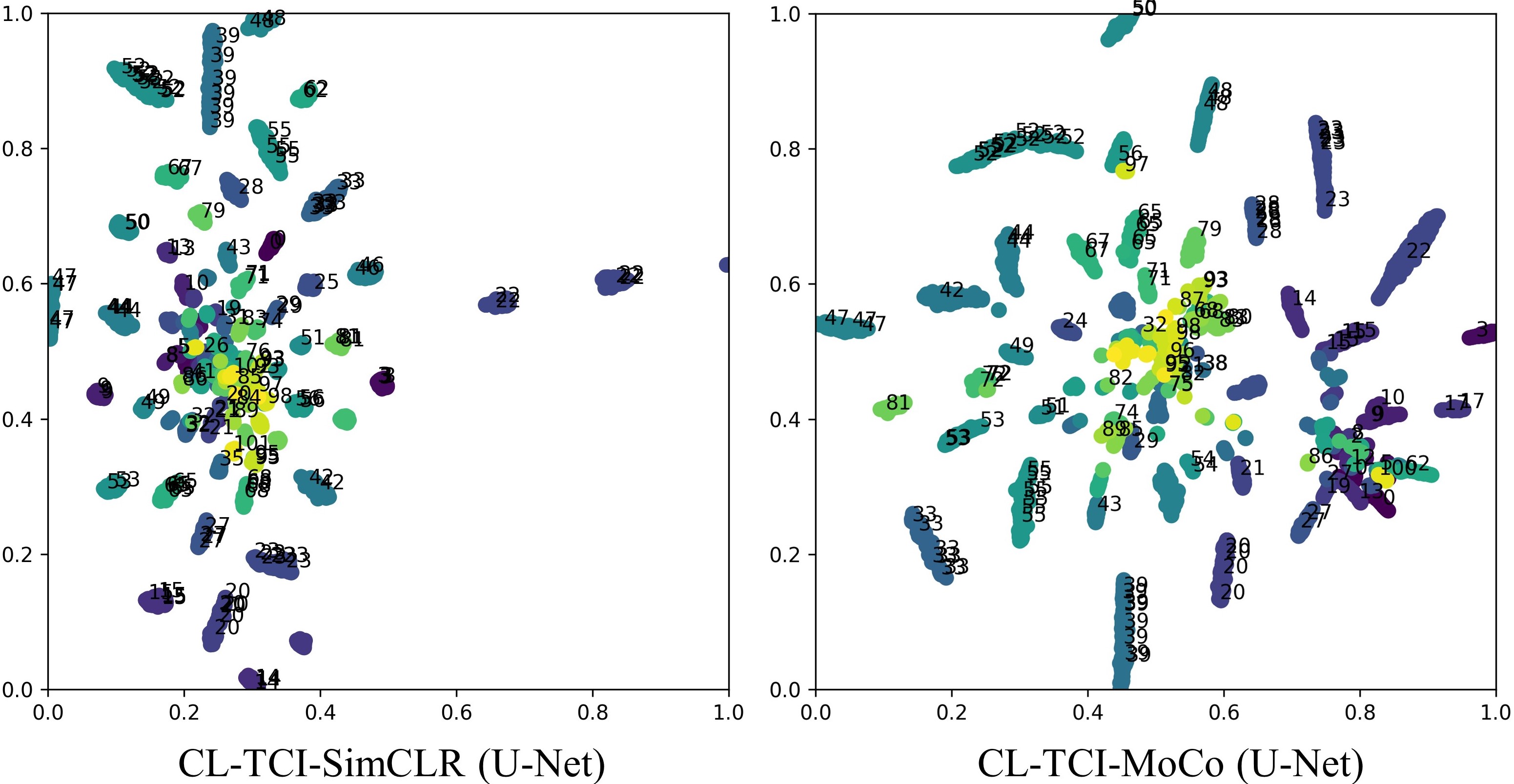}
\caption{T-SNE plots of CL-TCI-SimCLR and CL-TCI-MoCo.}
\vspace{-12pt}
\label{fig:tsne}
\end{figure}

In this section, we use the t-SNE to visualize the representations generated by our pre-trained model using data from the BCH-CXR dataset.
The t-SNE plots of CL-TCI-SimCLR and CL-TCI-MoCo using U-Net backbone can be seen in Fig. \ref{fig:tsne}.
Each dot in the plot represents one CXR.
The number beside each dot is the unique patient id, CXRs of the same patient are marked with the same color.
From the plots, we can see that CXRs of the same patients are well clustered, which means the contrastive learning has helped the network learn features to distinguish CXRs from different patients.
These learned features are highly likely to improve the performance of downstream segmentation tasks as we have seen in the quantitative results.

\section{Conclusion}

In this paper, we propose a novel contrastive learning framework called CL-TCI to leverage the information of temporal correlated medical images for lung segmentation tasks in CXR.
We adapt our CL-TCI from two state-of-the-art contrastive learning methods called SimCLR and MoCo and found CL-TCI-MoCo performs the best because it can utilize the contrastive pairs more effectively.
Experiment results on our own collected BCH-CXR dataset and two public available datasets demonstrate our CL-TCI framework outperforms all the baselines in both semi-supervised settings and transfer learning settings.

\noindent \textbf{Acknowledgements.} The study was approved by the institutional review board of Boston Children’s Hospital (BCH) (IRB-P00037536) under exemption from informed consent.  All chest radiographs for a series of consecutive cardiac patients receiving a chest radiograph in June 2021 were identified, anonymized, and exported for review. 

\bibliographystyle{IEEEtran}
\bibliography{references.bib}

\end{document}